\documentclass{PoS}
\usepackage{epstopdf}

 \def\thjet{\theta_{\rm j}}

\title{The Swift short gamma-ray burst rate density: prospects for detecting binary neutron star mergers by aLIGO}

\ShortTitle{Prospects for detecting binary neutron star mergers by aLIGO}

\author{\speaker{David M. Coward}$^{1,2}$, 
Eric Howell$^{1}$,
Tsvi Piran$^{3}$, Giulia Stratta$^{4}$, Marica Branchesi$^{5,6}$, Omer Bromberg$^{3}$, Bruce Gendre$^{4,8}$, Ronald Burman$^{1}$ and Dafne Guetta$^{7,8}$\\
 $^{1}$School of Physics, University of Western Australia, Crawley WA 6009, Australia\\
$^{2}$Australian Research Council Future Fellow\\
$^{3}$Racah Institute of Physics, The Hebrew University, Jerusalem 91904, Israel\\
$^{4}$ASI Science Data Center, via Galileo Galilei, 00044 Frascati (RM), Italy\\
$^{5}$DiSBeF - Universit\`a degli Studi di Urbino `Carlo Bo', I-61029 Urbino, Italy\\
$^{6}$INFN, Sezione di Firenze, I-50019 Sesto Fiorentino, Italy\\
$^{7}$Department of Physics and Optical Engineering, ORT Braude, P.O. Box 78, Karmiel, Israel\\
$^{8}$ INAF - Osservatorio Astronomico di Roma, Via Frascati 33, I-00040 Monteporzio Catone (Roma), Italy\\
 E-mail: \email{david.coward@uwa.edu.au}, \email{eric.howell@uwa.edu.au}, \email{tsvi@phys.huji.ac.il}, \email{stratta@asdc.asi.it}, \email{marica.branchesi@uniurb.it}, \email{omer@wise.tau.ac.il}, \email{bruce.gendre@asdc.asi.it}, \email{ron.burman@uwa.edu.au}, \email{dafne.guetta@oa-roma.inaf.it}}

\abstract{Presently only 30\% of short gamma ray bursts (SGRBs) have accurate redshifts, and this sample is highly biased by the limited sensitivity of {\it Swift} to detect SGRBs.
We account for the dominant biases to calculate a realistic SGRB rate density out to $z = 0.5$ using the {\it Swift} sample of peak fluxes, redshifts, and those SGRBs with a beaming angle constraint from X-ray/optical observations. Assuming a significant fraction of binary neutron star mergers produce SGRBs, we calculate lower and upper detection rate limits of (1-180) per Yr by an advanced LIGO and Virgo coincidence search. Our detection rate is compatible with extrapolations using Galactic pulsar observations and population synthesis.
}

\FullConference{Gamma-Ray Bursts 2012 Conference -GRB2012,\\
		May 07-11, 2012\\
		Munich, Germany}

\begin{document}

\section{Introduction}
Binary neutron star mergers (NS-NS) are favoured as the progenitors for short gamma ray bursts (SGRBs), based on the association of some SGRBs with an older stellar population, and
host galaxy types \cite{2005ApJ...630L.165L,2007ApJ...665.1220Z}. Kicks imparted to NSs at birth will produce velocities of several hundred km s$^{-1}$, implying that binary inspiraling systems may
occur far from their site of origin. Fong, Berger \& Fox \cite{2010ApJ...708....9F} using Hubble Space Telescope observations to measure SGRB-galaxy offsets, find the offset distribution compares favourably with the predicted distribution for NS-NS binaries.

Within 5 years, co-ordinated gamma-ray, X-ray, optical and gravitational-wave observations may allow the strong gravity regime of the central engine of compact object mergers to be probed. Such `multi-messenger' observations provide the opportunity to probe these events across a vast energy spectrum, and to constrain the progenitor populations of SGRBs. Furthermore, co-ordinated optical and gravitational-wave searches may play an important role in confirming the first direct gravitational-wave observations of compact object mergers \cite{2011MNRAS.415L..26C}. It is becoming increasingly important to constrain the rate of compact object mergers and their proposed optical counterparts in the context of up-coming gravitational-wave searches.

We calculate a beaming-corrected SGRB rate density using the {\it Swift} sample of SGRB peak fluxes, redshifts and inferred beaming angles from X-ray observations.
For our GRB selection criteria we use the Jochen Greiner catalogue of localized GRBs (see Table 1) and select bursts indicated as short that have reliable redshifts up to 2012 April.
We avoid using a SGRB luminosity function, models for progenitor rate
evolution, and a beaming angle distribution, all of which have large uncertainties. Instead, we focus on observed and
measured parameters that take into account selection effects
that modify {\it Swift}'{\it s} detection sensitivity to SGRBs. Finally, we use our SGRB rate density estimates to infer a detection rate of binary NS mergers by advanced LIGO (aLIGO) and Virgo (AdV) interferometers. Despite the poor
statistics, this approach gives meaningful results and
can be followed up when a larger sample of SGRB observations becomes available.

\section{Model}
The intrinsic SGRB rate is calculated taking into account the
following observational bias effects using the sample in
Table 1. GRB051221A is the only SGRB with a directly measured beaming
angle in the sample (only a lower limit for the beaming angle was obtained for GRB 070724A--see Coward et al. \cite{mnrat12}). We evaluated an upper SGRB rate density by using the
smallest observed beaming angle in the sample (7 deg) and a lower rate by assuming isotropic emission.
\begin{table}
 \begin{tabular}{@{}lcccc}
\hline
\hline
SGRB  & $T_{90}$ &  $\thjet$   &  $z$   &  peak flux \\
      & (s)  & (deg) &  & (ph s$^{-1}$ cm$^{-2}$) \\
\hline
\hline
101219A$*$  & 0.6 &  - &  0.718  & 4.1   \\
100117A$*$  & 0.3 &   - &  0.92  & 2.9   \\
090510A$*$   & 0.3 &   - &  0.903  & 9.7   \\
080905A\boldmath {$^a$}  & 1.0 &   - &  0.122  & 6.0    \\
070724A  & 0.4 &   $>11$$^\dagger$ &  0.457  & 2.0    \\
061217A   & 0.2 &   - &  0.827  & 2.4   \\
051221A$*$ & 1.4  &   7 &  0.547  & 12.0   \\
050509B  & 0.73 &   - &  0.225  & 3.7    \\
\hline
\hline

\end{tabular}
\caption{SGRB peak fluxes, $T_{90}$ and redshifts taken from the \emph{Swift} online catalogue and http://www.mpe.mpg.de/$\sim$jcg/grbgen.html used to calculate Poisson rates. We use the 20-ms peak photon fluxes from the BAT2 catalogue where possible -- those marked by $*$ are 1-s peak photon fluxes.
$^\dagger$ We set a lower limit on the jet opening angle from the time when the {\it Swift} XRT monitoring stopped.{\boldmath $^a$}The proposed host galaxy at $z = 0.1218$ for GRB 080905A \cite{2010MNRAS.408..383R} is a strong outlier to the Yonetoku relation \cite{2004ApJ...609..935Y} and a redshift $z > 0.8$ would make it consistent \cite{gruber12}. See \cite{mnrat12} for expanded table with references.}
\end{table}

The low energy detection bandwidth of {\it Swift} (15--150 keV) in comparison with BATSE's 50--300 keV results in a bias against SGRBs which typically have harder emissions. Secondly, the {\it Swift} detection threshold is not simply defined by the detector sensitivity, but by a complex triggering algorithm \cite{gehrels08}. Both these effects manifest as a bias against {\it Swift} detecting SGRBs; i.e. a smaller proportion of bursts has been detected by {\it Swift} ($\sim$10\%) \cite{gehrels09}. The latter effect results from the detection process employed by BAT, the \emph{Swift} coded-aperture mask $\gamma$-ray detector. In addition to requiring an increased photon count rate above background (the sole triggering criterion used for BATSE), BAT employs a second stage in which an image is formed by accumulating counts for up to 26\,s \cite{2006ApJ...644..378B}.

We attempt to crudely correct for these biases by using the observed SGRB rate from BATSE as a rate calibration for the \emph{Swift} SGRB rate. Because BATSE operated at different energy thresholds and trigger sensitivities \cite{2006ApJ...644..378B}, for consistency we take
all BATSE SGRBs with 64-ms peak flux when the trigger threshold was set to 5.5 $\sigma$ in the 50--300 keV energy range (total live operation time of 3.5 years). This yields 32 SGRBs sr$^{-1}$ yr$^{-1}$, assuming an effective BATSE FoV of $\pi$ sr \cite{2003ApJ...588..945B}. 

Because SGRBs occur over a short duration, it is more difficult (compared to long bursts) to produce a significant signal above background.  Hence instead of using the theoretical BAT sensitivity of $F_{\mathrm{Lim}}=0.4$ ph\,s$^{-1}$ cm$^{-2}$ \cite{2011MNRAS.415L..26C}, we employ a flux limit of 1.5 ph\,s$^{-1}$ cm$^{-2}$, using the smallest 20-ms peak flux from the SGRB data.

Taking into account the sensitivity reduction and k-correction (see \cite{mnrat12} for a derivation of the k-correction), the SGRB all-sky rate can be inferred from the flux-limited SGRB sample in Table 1. We calculate the maximum distance $d_{\mathrm{max}}$, with corresponding redshift $z_{\mathrm{max}}$, that a burst at luminosity distance $d_L$ could be detected given \emph{Swift's} sensitivity $F_{\mathrm{Lim}}$ (see \cite{mnrat12} for the derivation of $z_{\mathrm{max}}$).

The corresponding maximum SGRB detection volume for each burst is defined as
\begin {equation}
V_{\mathrm{max}} = \int_{0}^{z_{\mathrm{max}}} \frac{dV}{dz} dz\;,
\end{equation}
where the volume element, $dV/dz$, and luminosity distance, $d_L(z)$, are calculated using a flat-$\Lambda$ cosmology with $H_{\mathrm 0}$ = 71 km s$^{-1}$ Mpc$^{-1}$, $\Omega_M$ = 0.3 and $\Omega_\Lambda$ = 0.7.

To calculate the intrinsic rate of SGRBs requires accounting for the beaming angle, $\thjet$, of the jetted burst. Equation (\ref{eq3}) expresses the beaming factor used to correct for the unobserved SGRBs that are not detected because the jet is misaligned with the detector:
\begin{equation}
B(\thjet) =  [1- \mathrm{cos} (\thjet)]^{-1}\,.
\label{eq3}
\end{equation}

To account for the fact that only a fraction of observed SGRB have measured redshifts, we scale the rate density by the ratio of {\it Swift} bursts with redshift to those without redshifts, $F_r \approx 8/39$. The time span encompassing all observations, $T\sim6$ yrs, is defined by the start of {\it Swift} observations to the time of the most recent SGRB in the sample and we account for the fractional sky coverage of {\it Swift}, $\Omega\approx0.17$. To account for {\it Swift}'{\it s} reduced sensitivity for detecting SGRBs relative to BATSE, we use the ratio of the BATSE to {\it Swift} SGRB detection rate, which we approximate as $R_{B/S}=6.7$. This converts the {\it Swift} SGRB rate to an intrinsic SGRB rate. We point out that this correction applies only to SGRB (see \cite{mnrat12} for an analysis including SGRBs with extended emission). Finally, we use the probabilities, P$_{{i}(T_{90}; {\rm P}_{\rm L})}$, that each burst is a non-collapsar \cite{brom12,brom12b}, and scale the rate density by these probabilities. We note that these probabilities do not affect the rate densities significantly.

Combining all detection parameters yields the Poisson SGRB rate density of a single ($i$th) burst, and the total rate density for $n$ bursts:
\begin {equation}\label{Rate}
R_{\mathrm{SGRB}} = \sum_{i}^{n}  \frac{1}{V_{\mathrm{i(max)}}} \frac{1}{F_r} \frac{1}{T} \frac{1}{\Omega} R_{B/S}  B_i(\thjet) {\rm P}_{{i}(T_{90}; {\rm P}_{\rm L})}.
\end{equation}

\section{Results and summary}

\begin{table}
 \begin{tabular}{@{}lcccc}
\hline
\hline
      SGRB  & P$_{{i}(T_{90}; {\rm P}_{\rm L})}$  &     lower rate   &   upper rate  \\
         &   &  Gpc$^{-3}$ yr$^{-1}$             &  Gpc$^{-3}$ yr$^{-1}$ \\
\hline
\hline
101219A  &  0.79 & 0.039 &  5.3$^\dagger$\\
100117A  &  0.89 & 0.039 &  5.2$^\dagger$\\
090510   &  0.89 & 0.02 &  2.7$^\dagger$\\
080905A\boldmath {$^a$}  &  0.72 & 4.9 &  660$^\dagger$\\
070724A  &  0.84 & 0.77 &  140\\  
061217   &  0.89 & 0.2 &  27$^\dagger$\\
051221A  &  0.72 & 0.026 &  3.5\\
050509B  &  0.79 & 2 &  270$^\dagger$\\   

\hline
Total rate & &   $8^{+5}_{-3} $ & $1100^{+700}_{-470} $\\
\hline

\hline
\hline
\end{tabular}
\caption[]{The beaming-corrected SGRB rate densities with Poisson uncertainties using the observed constraints on $\thjet$, and scaled by the probability P$_{{i}(T_{90}; {\rm P}_{\rm L})}$, except GRB 100816A, which uses just the $T_{90}$.
Lower rate estimates assume isotropic emission, and upper rates use the observed beaming angle constraints shown in Table 1, or the smallest observed beaming angle in the sample$^\dagger$, $\thjet\approx7^o$. 
 \boldmath {$^a$} Because of the importance of GRB 080905 for the rate density, and its uncertainty (see Table 1. for caveats), we also calculate total rates excluding this burst i.e. $(3^{+2}_{-1}-500^{+340}_{-220})$ Gpc$^{-3}$ yr$^{-1}$.} 
\label{table2}
\end{table}

\begin{figure}
\includegraphics[scale=0.75]{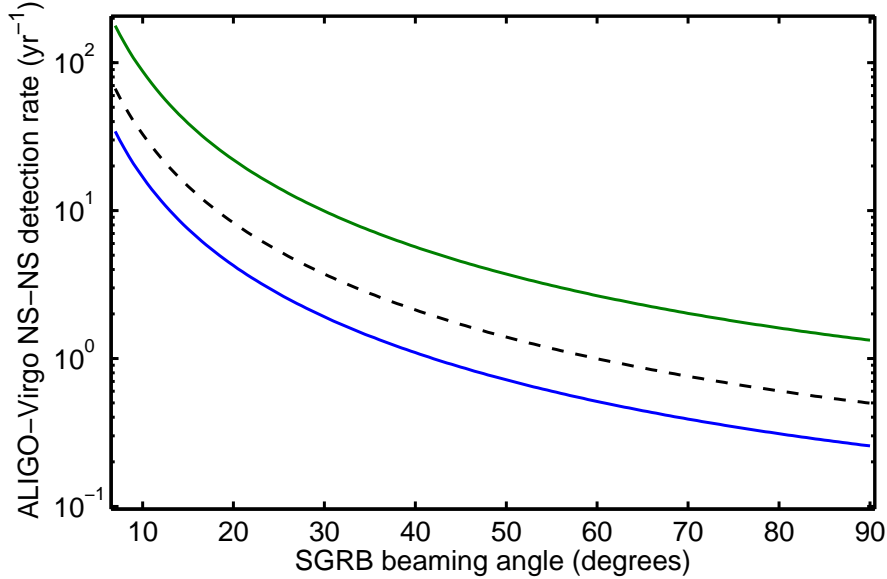}
\caption{The lower curve plots equation 3.1, the detection rate of binary NS mergers by a single aLIGO interferometer ($D_h=197$ Mpc) as a function of SGRB beaming angle, using $R_{\mathrm{Low}}=8$ $\mathrm{Gpc}^{-3}\mathrm{yr}^{-1}$ (see Table 2). The upper curve assumes a three detector coincidence search with $D_h=340$ Mpc. Both horizon distances used to calculate detection rates are angle averaged over all binary orientations. The dashed curve is the same as the upper curve but using the upper rate excluding GRB 080905 (see Table 1. for caveats).} \label{fig1}
\end{figure}

We use the above SGRB rate density to infer a detection rate of binary NS mergers by advanced gravitational-wave interferometers. For aLIGO and AdV interferometer sensitivities, the horizon distance, $D_h$ (all sky locations and orientation averaged over) for optimal detection of a NS-NS merger in a coincidence search is about 340 Mpc (cosmological redshift not included).  For a direct comparison with \cite{lscrate}, the detection rate is computed for a single interferometer, with $D_h=197$ Mpc, and for optimal detection in a three detector
coincidence search (see figure \ref{fig1}).  
Given the uncertainty in the beaming angle distribution, we define a detection rate as a function of $\thjet$ using the SGRB lower rate estimate, i.e. from $\thjet=90^o$, scaled by $B(\thjet)$ and the Euclidean volume: 
\begin{equation}
R(\thjet) = \frac{4\pi}{3} D_h^3 R_{\mathrm{Low}}B(\thjet).
\label{gw}
\end{equation}
For a realistic SGRB beaming angle range $(\thjet=7-30)$ degrees, the corresponding detection rate by a coincidence search can range from $(10-180)$ detections per year. 

The binary NS gravitational-wave detection rate estimates are based on calculating an intrinsic SGRB rate density using {\it Swift} localized bursts, taking into account dominant selection effects. This approach, based on observational data is very different from that based on Galactic binary pulsar observations and modelled population synthesis.
In the latter, \cite{lscrate} use the observed Galactic binary pulsar population 
and extrapolate a NS merger rate density out to the aLIGO and
AdV detection horizon. Conversely, our approach avoids this
extrapolation because it is essentially an observed rate extending
well beyond the average sensitivity distances of the upcoming
gravitational-wave searches for compact binaries (about 300 Mpc or $z =
0.07$ for aLIGO and AdV interferometers). 

In conclusion, the upcoming gravitational-wave detection era will be fundamental for resolving the SGRB--binary NS merger connection, since an unequivocal association between SGRBs and binary NS mergers will only be possible via coincident gravitational-wave and electromagnetic observations. Ultimately, a comparison beteween the SGRB rate density and the gravitational-wave detection rate will help constrain the fraction of binary NS mergers that give rise to SGRBs and the SGRB beaming angle distribution.

\acknowledgments
D. Coward thanks the organisers of GRB12 Munich for the opportunity to present this work, and for important feedback from the conference participants and referee.

\end{document}